\documentclass[journal]{IEEEtran}

\usepackage{amsmath,amsfonts}
\usepackage{algorithmic}
\usepackage{algorithm}
\usepackage{array}
\usepackage[caption=false,font=normalsize,labelfont=sf,textfont=sf]{subfig}
\usepackage{textcomp}
\usepackage{stfloats}
\usepackage{url}
\usepackage{verbatim}
\usepackage{graphicx}
\usepackage{cite}
\usepackage{xcolor} 
\usepackage{CJKutf8}
\usepackage{hyperref}
\usepackage{colortbl}
\usepackage{makecell}
\usepackage{amsmath}
\usepackage{amssymb}
\usepackage{graphicx}
\usepackage{pifont}
\usepackage{afterpage}
\usepackage{xspace}

\newcommand{\crossAttentionModule}{Dual Modal Enhancement Module}
\newcommand{\FusionModule}{Mutual Modal Aggregation Module}

\newcommand{\nName}{\emph{SpecSAR-Former}\xspace}
\newcommand{\dName}{\emph{Dynamic World+}\xspace}

\begin{document}

\title{\nName: A Lightweight Transformer-based Network for Global LULC Mapping Using Integrated Sentinel-1 and Sentinel-2}
﻿
\author{Hao Yu, \textit{Student Member, IEEE}, Gen Li, \textit{Student Member, IEEE}, Haoyu Liu, \textit{Student Member, IEEE}, Songyan Zhu, Wenquan Dong, Changjian Li, \textit{Member, IEEE}      % <-this % stops a space        % <-this % stops a space
\thanks{\fontfamily{ptm}\selectfont Hao Yu and Gen Li are co-first authors; Corresponding authors: Wenquan.dong@ed.ac.uk and Changjian.li@ed.ac.uk}. % <-this % stops a space

\thanks{\fontfamily{ptm}\selectfont Hao Yu is with the Agile Tomography Group, School of Engineering, the Institute for Imaging, Data and Communications, The University of Edinburgh,
Edinburgh EH9 3JL, UK.}
\thanks{\fontfamily{ptm}\selectfont Gen Li, Changjian Li are with the School of Informatics, University of Edinburgh, Edinburgh, UK.} % <-this % stops a space
\thanks{\fontfamily{ptm}\selectfont Haoyu Liu is with Tencent, Shenzhen, 518054, China.} % <-this % stops a space
\thanks{\fontfamily{ptm}\selectfont Songyan Zhu and Wenquan Dong are with the School of Geosciences, University of Edinburgh, Edinburgh EH9 3FF, UK.} % <-this % stops a space
}

\markboth{}%IEEE Transactions on Geoscience and Remote Sensing, ~Vol.~XX, No.~X, August~XXXX
{Shell \MakeLowercase{\textit{et al.}}: A Sample Article Using IEEEtran.cls for IEEE Journals}

% % column for its text to clear the IEEEpubid mark.

\maketitle 

\begin{abstract}

% ========= current reserach state =========
Recent approaches in remote sensing have increasingly focused on multimodal data, driven by the growing availability of diverse earth observation datasets. Integrating complementary information from different modalities has shown substantial potential in enhancing semantic understanding. 
% ========= motivation on proposed dataset =========
However, existing global multimodal datasets often lack the inclusion of Synthetic Aperture Radar (SAR) data, which excels at capturing texture and structural details. 
% ========= Add or not =========
SAR, as a complementary perspective to other modalities, facilitates the utilization of spatial information for global land use and land cover (LULC). 
% ========= our contribution on data =========
To address this gap, we introduce the \dName dataset, expanding the current authoritative multispectral dataset, Dynamic World, with aligned SAR data. 
% ========= our contribution on model =========
Additionally, to facilitate the combination of multispectral and SAR data, we propose a lightweight transformer architecture termed \nName. It incorporates two innovative modules, \crossAttentionModule{} (DMEM) and \FusionModule{} (MMAM), designed to exploit cross-information between the two modalities in a split-fusion manner.
These modules enhance the model's ability to integrate spectral and spatial information, thereby improving the overall performance of global LULC semantic segmentation.
Furthermore, we adopt an imbalanced parameter allocation strategy that assign parameters to different modalities based on their importance and information density.
% ========= experimental results =========
Extensive experiments demonstrate that our network outperforms existing transformer and CNN-based models, achieving a mean Intersection over Union (mIoU) of 59.58\%, an Overall Accuracy (OA) of 79.48\%, and an F1 Score of 71.68\% with only 26.70M parameters. The code will be available at \textit{\url{https://github.com/Reagan1311/LULC_segmentation}}

\end{abstract}


\section{Introduction}

% \todo{1) Refine figure and table captions; (Checked by hao) 2) Refine abstract; (Revised) 3) Figure 6, reorganize; (Revised) 4) Refine the method section; 5) Refine introduction (Revised)} 

% % \IEEEPARstart{T}{his} 
% \CJ{The introduction is too long. Is it the conversion of TGRS to have a long introduction?}
% \HY{P1 - Resived.}

% \CJ{P1 - LULC maps are important. The 'global' LULC seems like a unique feature, but did not emphsized.}
% \HY{P1 - Whether it is global or local LULC depends mainly on the dataset. Solution: When introducing the Dynamic World dataset in the Introduction section, it is mentioned that it is currently the Sota dataset of global LULC.}

% ========= Background =========
\IEEEPARstart{G}{lobal} land use and land cover (LULC) mapping is critical for understanding environmental changes, managing natural resources, and supporting sustainable development, providing essential data support for urban planning, agriculture, forestry, and conservation efforts \cite{foley2005global, gong2013finer, wang2023high}. As the Earth's surface undergoes rapid changes due to both natural processes and human activities, there is an increasing demand for accurate and detailed LULC maps with high spatial and temporal resolution at the global scale \cite{winkler2021global, gottlieb2024evidence}.

% ========= multimodal deep learning-based methods using multimodal data (VIS DSM LiDar)=========
% \CJ{P2 - Existing learning-based methods obtained significant performance boost.}
Existing studies by \cite{zhao2023land, papoutsis2023benchmarking, sathyanarayanan2020multiclass} have demonstrated the significant enhancements that deep learning techniques bring to the accuracy and efficiency of LULC mapping. Among these, in recent years, owing to the enhancement of computational capabilities and the substantial upgrades in hardware resources, multimodal deep learning-based methods have demonstrated considerable success in remote sensing LULC tasks \cite{bergamasco2023dual, LI2022102926}. For instance, Ma et al. \cite{ma2024multilevel} utilized visible image (VIS) and digital surface model (DSM) data in conjunction with FTransUNet for LULC classification, while Roy et al. \cite{roy2023multimodal} integrated hyperspectral image (HSI) and Light Detection and Ranging (LiDAR) data by the adapting traditional transformer architecture for the same purpose.

% \CJ{P3 - New data modalities bring both opportunities and challenges, preliminary attempts did not solve the challenges well.}

% =========Why use Sentinel-1 and Sentinel-2=========

In addition to the aforementioned multimodal data, recent advancements in earth observation technologies, particularly the European Space Agency's Sentinel missions \cite{berger2012esa}, have provided Sentinel-1 and Sentinel-2 data. In contrast to the previously mentioned data sources, Sentinel data offer superior spatial and temporal resolution, enabling detailed LULC mapping at both regional and global scales. 
% ============= Detailed information about Sentinel-1 and Sentinel-2 ==================
Multispectral data from Sentinel-2, acquired across multiple spectral bands ranging from visible to shortwave infrared wavelengths, provides a wealth of spectral information that is highly valuable for land cover classification tasks. The distinctive spectral signatures exhibited by various land cover types can be effectively captured, enabling accurate discrimination and mapping of diverse land cover classes with unique spectral characteristics \cite{drusch2012sentinel}. Synthetic aperture radar (SAR) data from Sentinel-1, on the other hand, offers a complementary perspective by leveraging the unique properties of radar signals. SAR's sensitivity to geometric and structural properties of targets provides valuable information for delineating land cover classes with varying surface roughness, moisture content, and vertical structures \cite{ienco2019combining}. However, interpreting SAR data visually can be challenging and lacks the spectral diversity of optical data. 

% ============ Machine learning methods have some problems =====================
Integrating these complementary data offers great potential to enhance LULC classification accuracy and robustness. Steinhausen et al. \cite{steinhausen2018combining} utilized a random forest-based forward feature selection approach to identify the most relevant Sentinel-1 radar scenes, optimizing classification accuracy by combining these with Sentinel-2 optical data. Tavares et al. \cite{tavares2019integration} integrated Grey-Level Co-occurrence Matrix (GLCM) textures from Sentinel-1 and normalized radiometric indices from Sentinel-2 to improve classification accuracy. However, due to the failure of these machine learning methods to effectively extract data features and form robust classification mappings, there remains room for improvement in their performance on classification tasks.
% ============ Multimodal Deep learning methdods for Sentinel-1 and Sentinel-2 LULC  =====================
Multimodal deep learning methodologies have demonstrated their capability in efficiently extracting relevant information from Sentinel-1 and Sentinel-2 data for LULC mapping. For instance, Hosseiny et al. \cite{hosseiny2021wetnet} developed a deep learning-based model for wetland classification that integrates Sentinel-1 and Sentinel-2 data within a tripartite architectural framework, resulting in a concatenated feature space and an ensemble classification process through majority voting. Furthermore, Gbodjo et al. \cite{gbodjo2021multisensor} introduced a multi-branch CNN framework designed to incorporate multimodal data, specifically Sentinel-1 SITS and SPOT along with Sentinel-2 SITS, to enhance the accuracy of LULC mapping. 

\begin{figure}[!tb]
    \centering
    \includegraphics[scale=1]{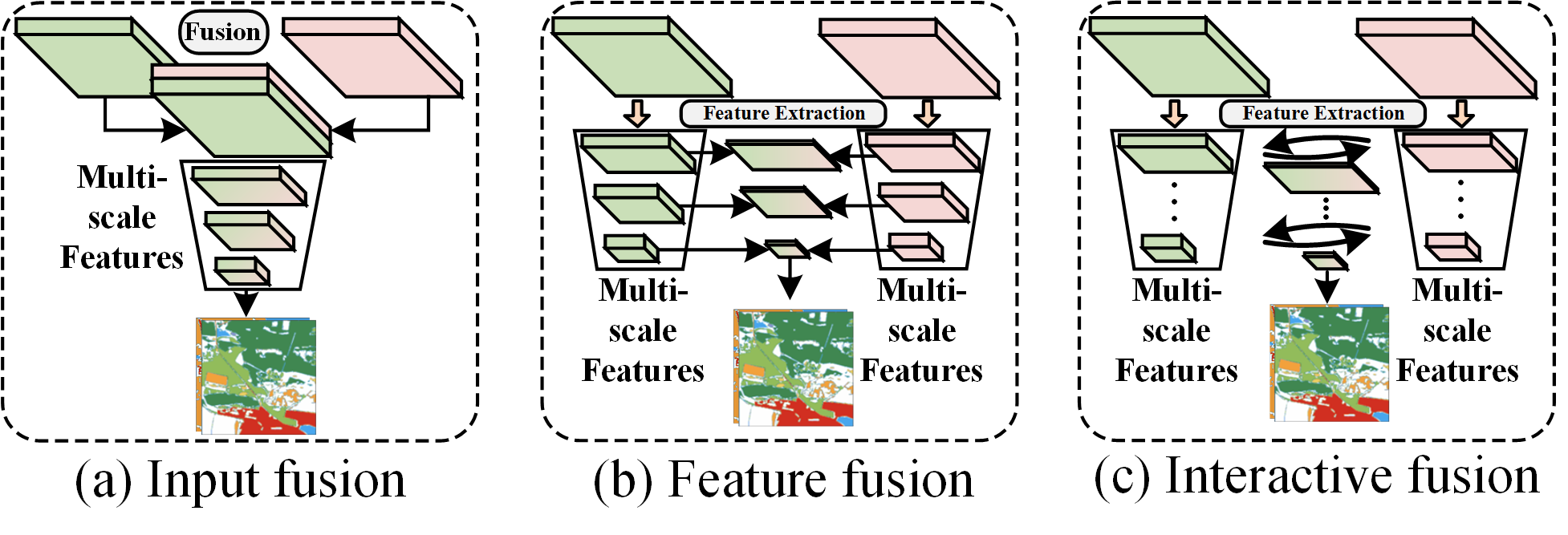}
    \caption{ Comparison of various fusion techniques. (a) Input fusion integrates inputs using modality-specific operations. (b) Feature fusion employs an attention-based module to merge features in a unidirectional manner. (c) Our interactive fusion method introduces bidirectional cross-modal feature rectification. \label{fig:Different_Fusion}}

\end{figure}

Although these studies are concerned with LULC tasks, they do not encompass global coverage. This limitation is partly due to the absence of comprehensive global datasets that include both multispectral and SAR data.  
% ============  Existing limitation (Network) =====================
Besides, although previous multimodal deep learning networks have shown promising performance, there are still limitations in feature representation: 1) The inherent imaging differences between SAR and spectral images result in varying contributions to semantic segmentation. When building feature extraction networks, the heterogeneity of these multisource data features is often overlooked, and training across different data sources with the same parameter settings. 2) Existing transformer-based models typically only linearly fuse multiscale information and fail to effectively utilize the interactions among multiscale information extracted from different modalities during the training process, which can lead to feature redundancy. 3) Some of the complex fusion methods implemented often incorporate a substantial number of parameters, potentially escalating the complexity of the network.

% \CJ{P5 - Our method.}
% ============   Our method =====================
To tackle these challenges, this paper first enhances the Dynamic World dataset by developing the \dName dataset, which synchronizes Sentinel-1 data, processed through a temporal sliding window, with Sentinel-2 data. Subsequently, this paper proposes a lightweight network, \nName. The network possesses hierarchical modal interaction Transformer encoders for efficient feature extraction and modal information interaction, and adopts a differentiated parameter configuration strategy that emphasizes key modalities.  The network consists of two branches: the multispectral perception branch primarily focuses on extracting spectral features from spectral data, while the spatial perception branch is dedicated to mapping the distribution relationships among different types of land cover and analyzing their textures and structures. Furthermore, to effectively facilitate the interaction and integration of multispectral and SAR data for global LULC mapping, we introduce a \crossAttentionModule{} (DMEM) and a \FusionModule{} (MMAM). Specifically, after the net extracts multi-scale information from modalities, the corresponding scale features achieve bidirectional interactive information sharing through the DMEM, as shown in Fig. \ref{fig:Different_Fusion}{}(c). These features are effectively fused in the MMAM, and after a splitting operation according to the modal importance, the processed features are passed on to the next cascaded encoder. This structural design significantly enhances the integration capability of the modality and scale information, thereby improving the model's accuracy and efficiency in global LULC mapping.

Extensive experiments demonstrate that \nName surpasses other state-of-the-art methods in global LULC semantic segmentation across mainstream benchmarks, while also excelling in terms of the number of parameters and FLOPs (refer to Fig. \ref{fig:miou_param_flops}). Additionally, \nName achieves a suitable inference speed. Ablation experiments further highlight the effectiveness of the individual components, input modalities, and parameter configuration strategy. The main contributions of the paper are summarized as follows:

\begin{figure}[!t]
    \centering
    \includegraphics[scale=1]{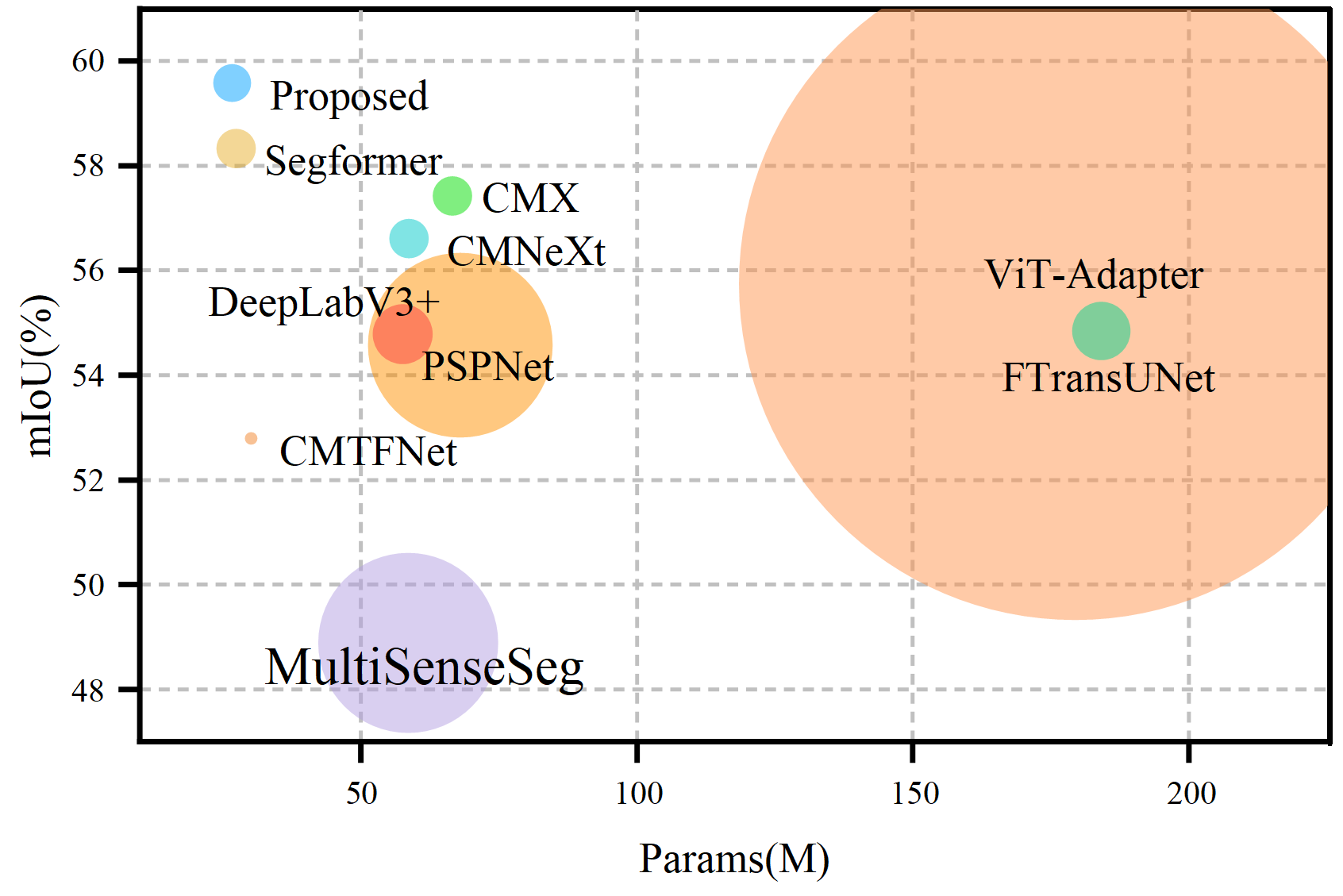}
    \caption{ Segmentation results from proposed \nName and other foundational models on the Google Dynamic World+ dataset, illustrated with bubbles where each bubble's size corresponds to the computational complexity (FLOPs) of the baseline models.
    \label{fig:miou_param_flops}}
\end{figure}

\begin{enumerate}

\item We expand the Dynamic World dataset to create \dName dataset, aligning Sentinel-2 data with Sentinel-1 data processed using a temporal sliding window for adaptability experiments.

\item We propose \nName, a lightweight network for multisource remote sensing semantic segmentation. It integrates spectral and structural features from multispectral and SAR data, respectively, enhancing long-range and fine-grained feature capture for global LULC classification. \nName excels in pixel-wise classification performance on the \dName benchmark, showcasing superior parameter efficiency and inference speed.

\item To better harness complementary information from diverse sources, we develop the DMEM and MMAM. These modules align cross-source features through multilevel information interaction and fusion, and a differentiated parameter configuration strategy according to modal importance. 
\end{enumerate}

The remainder of this article is organized as follows. Section II reviews previous studies on global LULC and multi-modal semantic segmentation methods based on Transformer. Section III introduces the extended \dName dataset. Section IV describes the architecture of the \nName, whereas Section V outlines the experimental setup. Sections VI discuss the experimental and ablation study results. The conclusion is provided in Section VII.

\section{Related Work}
\subsection{Global LULC Related Work}
In recent decades, significant efforts have been made to tackle the challenging task of global LULC. These efforts have resulted in a diverse range of LULC products with varying spatial resolutions and methodologies. Low-resolution products like UMD Land Cover \cite{hansen2000global} and Global Land Cover 2000 (GLC2000) \cite{bartholome2005glc2000} were among the early attempts. The resolution of these products is typically equal to or lower than 1000 m. Medium-resolution LULC products, such as GLCNMO (V2, V3) \cite{tateishi2014production} and MODIS/Terra+Aqua Land Cover Type (MCD12Q1) \cite{friedl2010modis}, offered 500 m spatial resolution based on the Moderate Resolution Imaging Spectroradiometer (MODIS) data. Products like ESA CCI Land cover \cite{cci2017product, bontemps2013consistent} and GlobCover \cite{arino2007globcover} used the Medium Resolution Imaging Spectrometer (MERIS) images with a 300 m resolution. These low-resolution land cover change products effectively capture broad spatial patterns of land cover types and quantify global land cover changes. However, they exhibit significant limitations, particularly in regions with intense human activity and high spatial heterogeneity. In such fragmented and heterogeneous landscapes, low-resolution satellite observations fail to accurately capture the detailed land cover changes.

Benefiting from the free availability of high-resolution satellite imagery and advancements in computational storage capabilities, high-resolution land cover products have experienced significant growth in recent years. These advancements have particularly impacted the development of LULC products with 30 m and 10 m resolutions. In 2013, the first global 30 m LULC product, FROM\_GLC, was introduced \cite{gong2013finer}. This was followed by other global LULC products based on Landsat data, such as GlobeLand30 \cite{chen2015global, chen2017towards}, GLC\_FCS30 \cite{zhang2020glc_fcs30} and GLC\_FCS30D \cite{zhang2024glc_fcs30d}. The European Space Agency's Copernicus program provided globally consistent optical and radar data from the Sentinel satellites. Based on this data, four products have been developed: Dynamic World \cite{brown2022dynamic}, ESA WorldCover \cite{zanaga2021esa, zanaga2022esa}, ESRI \cite{karra2021global}, and FROM-GLC10 \cite{gong2019stable}. These products exhibit two main limitations:

\begin{enumerate}

\item 
Only ESA WorldCover integrates both Sentinel-1 and Sentinel-2 data, while the other three products rely solely on multispectral data, not on multimodal data.

\item 
Additionally, only Dynamic World and ESRI utilize deep learning algorithms, but they implement relatively basic models; Dynamic World employs a Fully Convolutional Network (FCN), and ESRI utilizes a UNet.
\end{enumerate}

\subsection{Multi-modal Semantic Segmentation Methods based on Transformer}
Recent years have clearly marked a significant turning point for the use of transformer-based models in semantic segmentation. Transformer network, with its self-attention mechanism capable of capturing long-distance dependencies within data, not only effectively integrates information from various sources but also benefits from inherent support for parallel processing \cite{xu2021efficient,xiao2023enhancing}. These advantages make it particularly suited for complex multimodal tasks that rely on large datasets, such as global LULC semantic segmentation.

Among the newly proposed seminal works along this line, Xue et al. \cite{xue2022deep} developed a new deep hierarchical vision transformer (DHViT) architecture for the joint classification of hyperspectral and LiDAR data. Extended Vision Transformer (ExViT) \cite{yao2023extended} proposed by Yao et al. integrated separable convolution modules into position-shared ViTs to effectively handle multimodal remote sensing data, and employed a Cross-Modality Attention (CMA) module to enhance inter-modality information exchange, thereby improving classification accuracy. Roy et al. \cite{roy2024cross} developed a new multimodal deep learning framework by introducing a cross hyperspectral and LiDAR (Cross-HL) attention transformer to effectively fuse remote sensing data, aimed at improving LULC recognition. Jamali and Mahdianpari \cite{jamali2022swin} utilized the more advanced Swin transformer in conjunction with the VGGNet regional feature extractor for classifying complex RS landscapes. They employed Sentinel-1, Sentinel-2, and LiDAR data, which were suitable for tasks such as wetland mapping and biodiversity assessment.

\section{Dynamic World+ Dataset}

% \CJ{The overall feeling is that the necessity to have the two new subsets is not well delivered. Some details of the new subsets are not clear. The long introduction of Dynamic World dataset might be redundant.}

This study relies on labels from Dynamic World, which is currently the authoritative global 10m data proposed by Google~\cite{brown2022dynamic}. We processe the corresponding Sentinel-1 and Sentinel-2 data in Google Earth Engine. After data alignment, a total of 16,893 patches with both Sentinel-1 and Sentinel-2 coverage are used as training data, while 299 patches are used for testing and validation, which constitute the entire \dName dataset. Each patch consists of 510 pixels by 510 pixels.

\subsection{Dynamic World Label}
\begin{figure}[!tb]
  \centering
    \includegraphics[scale=1]{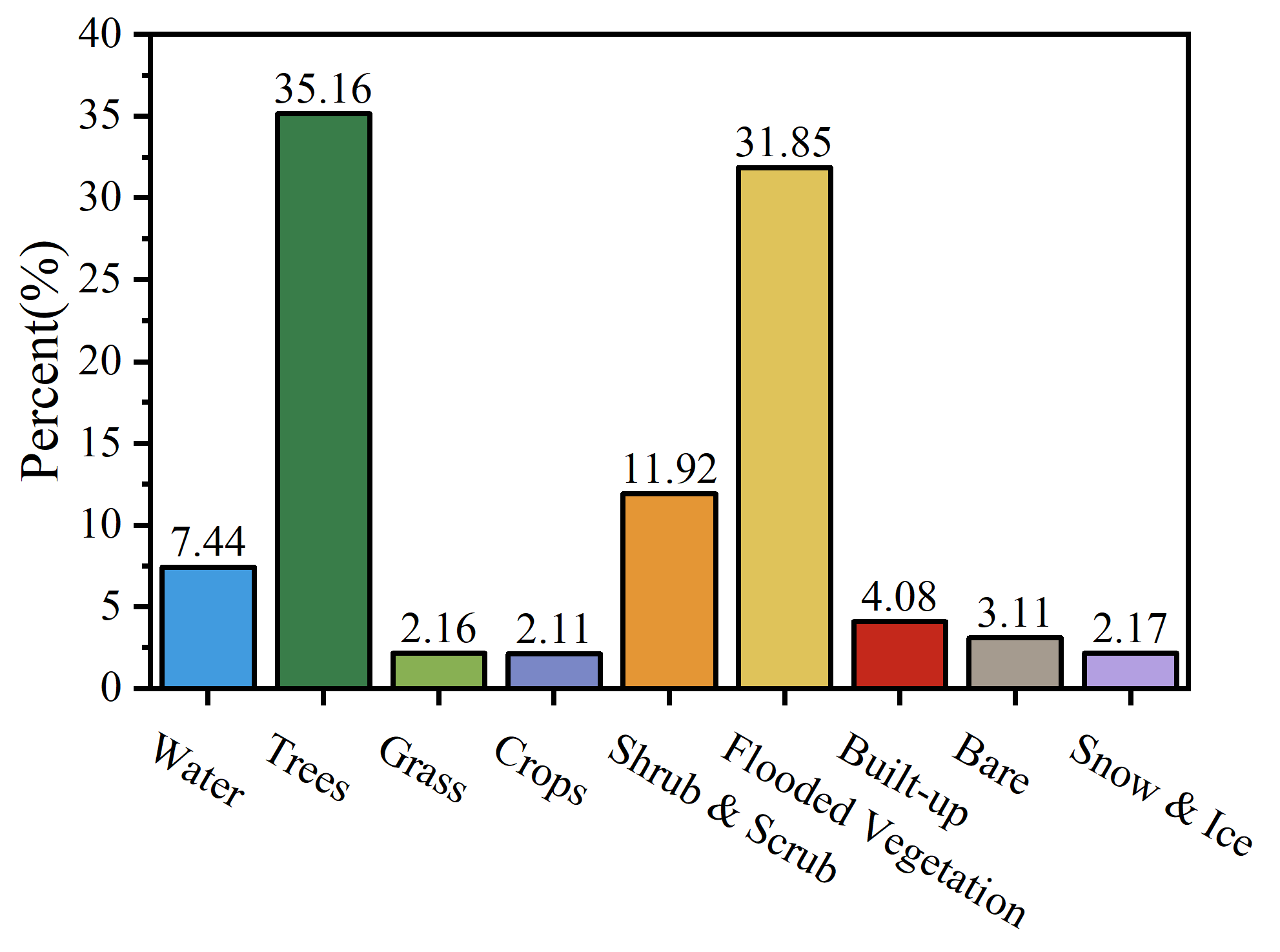}
    \caption{ Class distribution across 14 global biomes in the dataset.
 \label{fig:label}}
\end{figure}

The labeled data used in our study is sourced from Dynamic World, which encompasses an area exceeding 520,000 km\textsuperscript{2} globally. To the best of our knowledge, this represents the most extensive manually annotated global land use dataset currently available. Annotated by a team of 25 experts and 45 non-experts using Sentinel-2 imagery from 2019, the dataset classifies land cover into nine major categories: water, trees, grass, flooded vegetation, crops, scrub/shrub, built-up areas, bare ground, and snow/ice (Fig. \ref{fig:label}). The Dynamic World dataset employs a stratified sampling method across global regions and biomes, systematically dividing the Earth's surface into three major hemispheric regions, further subdivided according to established ecoregion classifications. This multi-tiered stratification approach ensures a geographically diverse and ecologically representative sampling framework. A cohort of three experts selects and annotates 409 image tiles, with the resulting expert consensus annotations used as validation data.

The breadth and diversity of this dataset, encompassing a wide range of global land cover types and vegetation patterns, provides a robust foundation for training models with applicability to global land cover classification tasks.

% \CJ{Is this well-known? What is done by us?}
% \CJ{Same for the labeling procedure. Is anything new done here?}

% ========================= Data labels =====================================
% ===========================================================================

\subsection{Sentinel-2}
Sentinel-2 is a high-resolution (10 m), multi-spectral imaging mission within the European Union's Copernicus Programme, operated by the European Space Agency (ESA) \cite{drusch2012sentinel}. The mission comprises a constellation of two identical satellites in the same orbit, phased at 180° to each other, providing a high revisit time of 5 days at the equator \cite{li2017global}.

The Sentinel-2 Multi-Spectral Instrument (MSI) captures data in 13 spectral bands, ranging from visible and near-infrared to shortwave infrared, at varying spatial resolutions (10m, 20m, and 60m) \cite{gascon2017copernicus}. We select 10 spectral bands from the Sentinel-2 data, specifically bands B1, B2, B3, B4, B5, B6, B7, B8, B8A, B11, and B12. 
% \CJ{Why? Jusfitication} 
All selected bands are resampled to a 10 m resolution using bilinear interpolation to ensure consistency with the spatial resolution of the label data.

In our study, we utilize Sentinel-2 Level-2A (L2A) data, which represents bottom-of-atmosphere reflectance in cartographic geometry \cite{main2017sen2cor}. The L2A data are acquired through Google Earth Engine, corresponding to the Dynamic World label dataset. Given that the cloud cover in these images had already undergone manual screening as part of the Dynamic World dataset preparation, we do not implement additional cloud removal procedures \cite{geudtner2014sentinel}.

\subsection{Sentinel-1}
Sentinel-1 is also a space mission within the European Union's Copernicus Programme, consisting of a constellation of two satellites equipped with C-band Synthetic Aperture Radar (SAR) instruments \cite{torres2012gmes}. This mission provides all-weather, day-and-night radar imaging for land and ocean services \cite{potin2019copernicus}.  Sentinel-1 offers a revisit time of 6 days at the equator with both satellites operational, and provides imagery at various spatial resolutions, ranging from 5 m to 100 m depending on the acquisition mode.

In our study, we use Sentinel-1 SAR Ground Range Detected (GRD) data in Interferometric Wide (IW) swath mode with a pixel size of 10 m. The IW mode is the main acquisition mode over land, providing a wide swath width of 250 km. The GRD products consist of focused SAR data that has been detected, multi-looked and projected to ground range using an Earth ellipsoid model \cite{filipponi2019sentinel}. To preserve the spatial information inherent in the SAR data, we leverage the temporal dimension of the data to reduce the speckle noise. Using Google Earth Engine, we retrieve all available SAR data within a 360-day window centered on the label date (180 days before and after). Some labels from Dynamic World do not have matching Sentinel-1 imagery within the time window, and these labels are discarded. We then compute the mean of these observations, effectively reducing speckle noise while maintaining spatial detail. The resulting VV (vertical transmit and vertical receive) and VH (vertical transmit and horizontal receive) band composites provide valuable information on surface roughness and structure, complementing the spectral information from optical sensors in our land cover classification efforts.

\section{METHODOLOGY}

% =============================== Network structure =========================
\begin{figure*}[!htb]
  \centering
    \includegraphics[width=1.8\columnwidth]{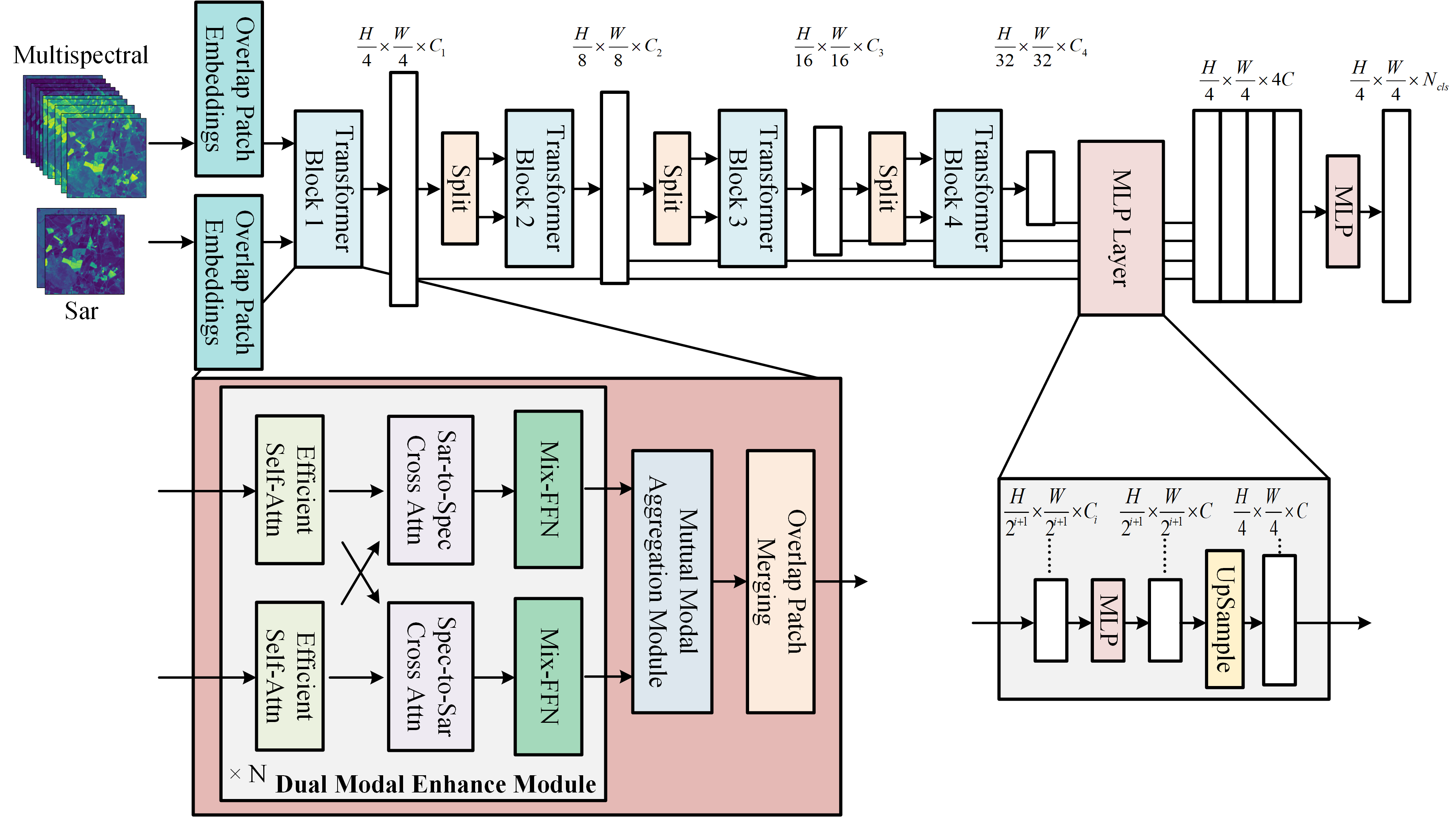}
    \caption{Schematic illustration of the proposed \nName. \label{fig:Network}}
\end{figure*}
% ===========================================================================

% \CJ{Overview section is too long and wordy. Should briefly introduce what is the overall design and insights.}
\subsection{Overall Architecture of Network}
The architecture of the proposed \nName, detailed in Fig. \ref{fig:Network} and drawing inspiration from Segformer \cite{xie2021segformer}, is specifically tailored and optimized for LULC semantic segmentation. To effectively process SAR and multispectral remote sensing data, the net comprises two branches: 1) A multispectral perception branch focuses on extracting features with unique spectral characteristics. 2) A spatial perception branch, which takes SAR as input, serves as the supplementary source of spectral information and is responsible for analyzing geometric and structural properties among different land cover types. Additionally, to enhance the interaction between two modal information and to facilitate the fusion of these modal features, this paper proposes the DMEM and MMAM, respectively. To ensure the focus of the network aligns with the significance of each data modality in global LULC mapping, this paper introduces a custom configuration strategy of branch parameters.

Specifically, given remote sensing data of size \(H \times W \times C_{\text{Spec}}\) and \(H \times W \times C_{\text{SAR}}\), we first segment them into \(4 \times 4\) patches, which facilitates dense prediction tasks. These patches are then input into a hierarchical modal interaction transformer encoder to extract multi-level features at resolutions of $1/4$, $1/8$, $1/16$, and $1/32$ of the original image resolution. The multi-level features obtained by the two branches first undergo corresponding efficient self-attention modules. After channel adjustments are made by the MLP, the process continues with SAR-to-Spectral and Spectral-to-SAR cross-attention operations to enhance the information flow and interaction between the multispectral and SAR features. Then, the features are operated by the corresponding Mix-Feed-Forward network (Mix-FFN) and fused using an MMAM, followed by an overlapped patch merging module. Subsequently, the features processed are split according to a $3/4$ and $1/4$ ratio, and these split features are then input into the next encoder. Lastly, the fused multi-level features are fed into the decoder, which predicts the semantic segmentation mask at a resolution of \( \frac{H}{4} \times \frac{W}{4} \times N_{\text{cls}} \), where \( N_{\text{cls}} \) is the number of categories.

\subsection{Hierarchical Modal Interaction Transformer Encoder}

\begin{figure}[!t]
  \centering
  \includegraphics[width=\linewidth]{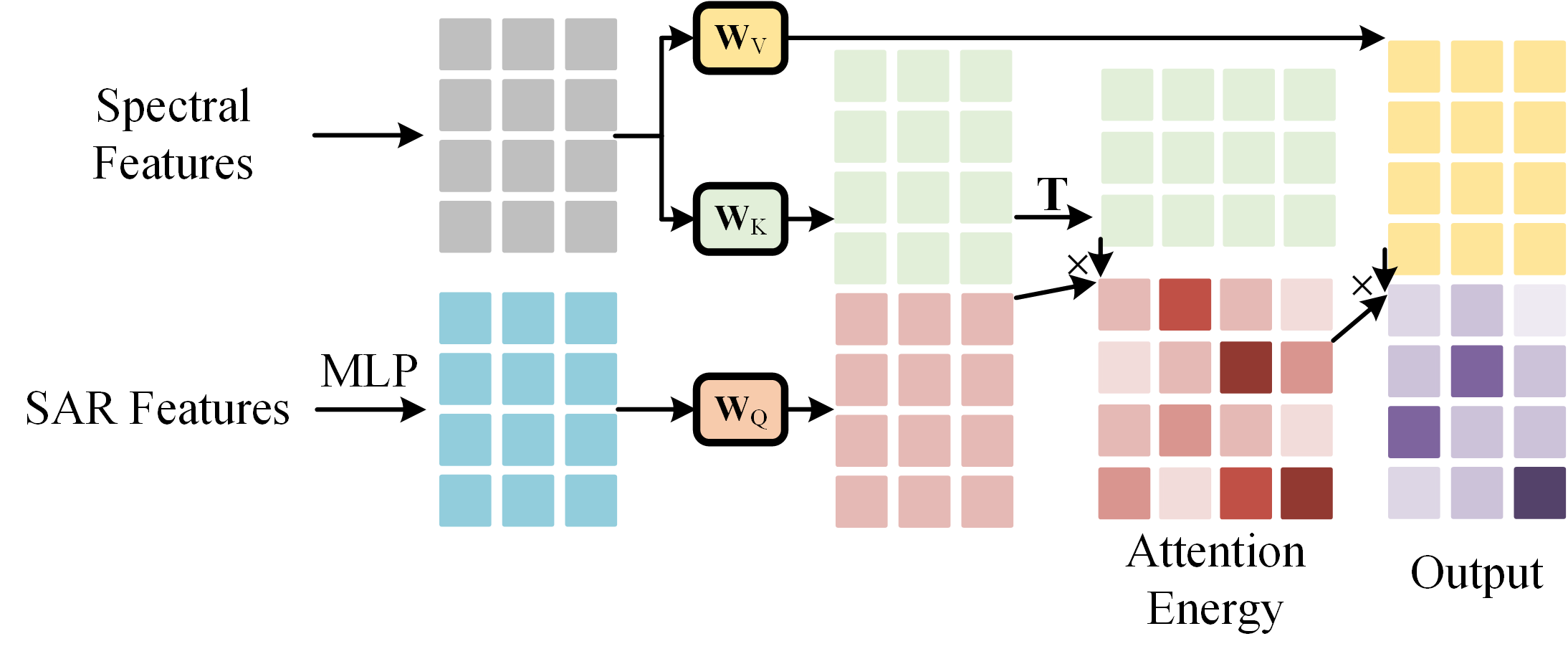}
  \caption{Illustration of the SAR-to-Spectral cross attention module.}
  \label{fig:CrossAtt}
\end{figure}

\begin{figure}[!t]
  \centering
  \includegraphics[width=\linewidth]{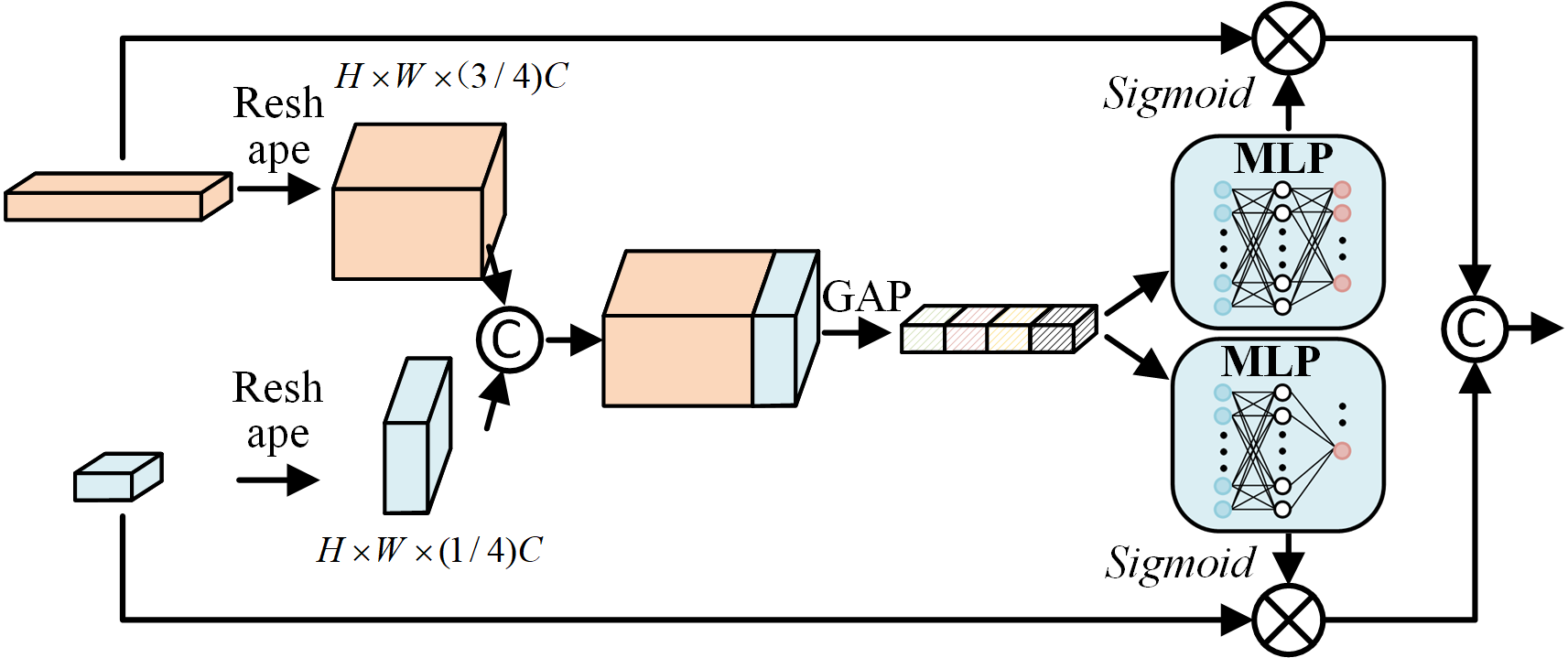}
  \caption{Illustration of the \FusionModule.}
  \label{fig:ChannelAtten}
\end{figure}

\subsubsection{Overlapped Patch Embeddings}
Different from the Vision Transformer (ViT), which employs a convolution with a kernel size equal to the stride size to produce non-overlapping patches, we follow~\cite{xie2021segformer} that generates hierarchical features via overlapped patch embeddings.
Specifically, the multispectral and SAR inputs $I_\text{Spec} \in\mathbb{R}^{H \times W \times C_\text{Spec}}$ and $I_\text{SAR} \in\mathbb{R}^{H \times W \times C_\text{SAR}}$ are first processed by a $7\times 7$ convolution kernel with stride 4 to get feature maps $F_1 \in\mathbb{R}^{\frac{H}{4} \times \frac{W}{4} \times C_{\text{Spec}_1}}$ and $F_1 \in\mathbb{R}^{\frac{H}{4} \times \frac{W}{4} \times C_{\text{SAR}_1}}$, respectively. 

% Utilizing the patch merging technique within the Vision Transformer (ViT) \cite{dosovitskiy2020image}, we effectively compress the dimensional space of our hierarchical features from $F_\text{Spec} (H \times W \times C_\text{Spec})$ and $F_\text{SAR} (H \times W \times C_\text{SAR})$ to $F_1 \left(\frac{H}{4} \times \frac{W}{4} \times C_{\text{Spec}_1}\right)$ and $F_1 \left(\frac{H}{4} \times \frac{W}{4} \times C_{\text{SAR}_1}\right)$, respectively. 

This process is then applied iteratively across the feature hierarchy via a $3\times 3$ convolution with stride 2.
The overlapping patch merging strategy enhances the efficiency of feature extraction and processing, and can maintain local continuity within the patches, which is crucial for preserving the integrity of spatial information and beneficial to the LULC senmantic segmentation.
\subsubsection{\crossAttentionModule} The DMEM consists of the efficient self-attention module, the SAR-to-Spectral cross attention module, the Spectral-to-SAR cross attention module, and the Mix-FFN module.\paragraph{Efficient Self-Attention} The traditional multi-head self-attention mechanism exhibits a computational complexity of \(O(N^2)\), which becomes impractical for processing large images. To mitigate the computational demands associated with dual modalities, a sequence reduction strategy \cite{wang2021pyramid} is employed, utilizing a reduction ratio \(R\):

\begin{equation}
\hat{K} = \text{Reshape}\left(\frac{N}{R}, C \cdot R\right)(K)
\end{equation}
\begin{equation}
K = \text{Linear}(C \cdot R, C)(\hat{K})
\end{equation}
where ${K}$ is the key vector of multi-head attention, and \( N = H \times W \) represents the sequence length. The Reshape function transforms $K$ into a format of $\frac{N}{R} \times (C \cdot R)$, and Linear $(C_{\text{in}}, C_{\text{out}})(\cdot)$ transforms a $C_{\text{in}}$-dimensional tensor into a $C_{\text{out}}$-dimensional tensor. This approach scales down the dimension of $K$ to $\frac{N}{R} \times C$, effectively reducing the self-attention complexity to $O\left(\frac{N^2}{R}\right)$.

\paragraph{SAR-to-Spec or Spec-to-SAR Cross Attention module} The SAR-to-Spec and Spec-to-SAR cross attention modules are specifically designed to handle two types of inputs: a primary input and an auxiliary input. For example, Fig. \ref{fig:CrossAtt} illustrates the SAR-to-Spec cross attention module, and the structure of the Spec-to-SAR cross attention module are similar. This module facilitates the injection of auxiliary priors into the primary modal branch, enabling the auxiliary branch to effectively provide complementary perspectives related to semantic segmentation for the primary branch, thereby enhancing the model's performance and efficiency. In the upper branch, the primary input consists of multispectral information, while in the lower branch, it processes SAR information.

The module generates query vectors by applying a linear transformation to the primary input. Initially, the dimension of the auxiliary input is adjusted through an MLP. The module then performs convolutional downsampling on the adjusted auxiliary input to reduce its dimensions, followed by layer normalization. Subsequently, another linear transformation is applied to the processed auxiliary input to produce key and value vectors. These key vectors are combined with the query vectors through scaled dot-product attention computations to generate attention scores, which are normalized using the Softmax function. Finally, these scores are used to weight and synthesize the final output features. Specifically, for the $i$-th block of the encoder, the principal feature $F^i_{\text{Enc}} \in \mathbb{R}^{H \times W \times C_{\text{Prin}}}$ is taken as the query, and the mapped complementary feature $F^i_{\text{Cf}} \in \mathbb{R}^{H \times W \times C_{\text{Comp}}}$ is taken as the key and value. Cross-attention is utilized to inject the complementary feature $F^i_{\text{Cf}}$ into the principal feature $F^i_{\text{Enc}}$, which can be formulated as follows:

\begin{equation}
\tilde{F}^i_{\text{Enc}} = F^i_{\text{Enc}} + \text{Attention}(\text{norm}(F^i_{\text{Enc}}), \text{MLP}(F^i_{\text{Cf}}))
\end{equation}
where $\text{norm}(\cdot)$ denotes Layer Normalization ($\text{LayerNorm}$), the attention layer $\text{Attention}(\cdot)$ employs sparse attention, and $\text{MLP}(\cdot)$ is used for dimensional adjustment.

\paragraph{Mix-FFN} Mix-FFN is introduced, leveraging zero padding's capability to transmit positional information, thus eliminating the need for traditional positional encoding in global LULC semantic segmentation tasks. This method integrates a $3\times3$ convolution directly within the FFN, streamlining the architecture and enhancing the model's efficiency. The formulation of Mix-FFN is as follows:

\begin{equation}
\mathbf{x}_{\text {out }}=\operatorname{MLP}\left(\operatorname{GELU}\left(\operatorname{Conv}_{3 \times 3}\left(\operatorname{MLP}\left(\mathbf{x}_{\text {in }}\right)\right)\right)\right)+\mathbf{x}_{\text {in }}
\end{equation}
where $x_{\text{in}}$ represents the feature vector obtained from the cross-attention module. This design integrates a \( 3 \times 3 \) convolution and an MLP in each FFN. Furthermore, we employ depth-wise convolutions to decrease the model's parameter count and enhance its computational efficiency.

\subsubsection{\FusionModule}
The MMAM processes and fuses two input features through global average pooling (GAP) and two parallel sequences of fully connected (FC) layers as shown in Fig. \ref{fig:ChannelAtten}. The module begins by concatenating the input features $F^i_{\text{Enc\_Spec}} \in \mathbb{R}^{H \times W \times \frac{3}{4}C}$ and $F^i_{\text{Enc\_SAR}} \in \mathbb{R}^{H \times W \times \frac{1}{4}C}$ along the channel dimension, followed by compressing the concatenated features globally using GAP. Subsequently, two parallel sequences of FC layers are used to extract and output two distinct sets of channel weights from the globally compressed information. These weights are subsequently processed through a sigmoid activation function and then applied to modulate the original input features, thereby enhancing significant features and attenuating less relevant ones. Finally, the weighted features are concatenated again to form the final output $F^i_{\text{Enc\_Out}} \in \mathbb{R}^{H \times W \times C}$. This structure effectively facilitates information fusion between features, enhancing the model's focus on significant characteristics. The formulation of the module is presented as follows.
\begin{equation}
{F}^i_{\text{Enc\_Out}} = \text{Concat}((M_1 \odot {F}^i_{\text{Enc\_Spec}}) , \text(M_2 \odot {F}^i_{\text{Enc\_SAR}}))
\end{equation}
\begin{equation}
M_1 = \text{Sigmoid}(\text{MLP}_1(\text{GAP}(\text{Concat}({F}^i_{\text{Enc\_Spec}}, {F}^i_{\text{Enc\_SAR}}))))
\end{equation}
\begin{equation}
M_2 = \text{Sigmoid}(\text{MLP}_2(\text{GAP}(\text{Concat}({F}^i_{\text{Enc\_Spec}}, {F}^i_{\text{Enc\_SAR}}))))
\end{equation}
where $\odot$ denotes the Hadamard product, and GAP stands for the global average pooling operation. Furthermore, $\text{MLP}_1$ and $\text{MLP}_2$ represent specific MLP operations.

\subsection{{Decoder}}
The proposed network utilizes a simplified decoder design, exclusively composed of MLP layers, eliminating the complex and resource-demanding components typically employed in conventional models. The all-MLP decoder architecture is constructed in four principal stages. Initially, multi-scale feature maps obtained after the hierarchical modal interaction Transformer encoder, denoted as \(F_i\), are processed through an MLP to standardize their channel dimensions. Following this, these features are upsampled to a quarter of their original resolution and subsequently aggregated via concatenation. The third stage involves the application of an additional MLP to integrate these concatenated features, denoted as \(F\), to ensure cohesive feature fusion. The final stage entails employing yet another MLP, which transforms the integrated features into the segmentation mask, represented as \(M\). This mask is structured to predict segmented regions at a reduced resolution of \(\frac{H}{4} \times \frac{W}{4} \times N_{\text{cls}}\), where \(N_{\text{cls}}\) signifies the number of global LULC categories. 

This design is aimed at enhancing the decoder's ability to generalize across various scales of semantic information after merging information from two branches, which is critical for precise semantic segmentation. The formulation of the decoder is as follows:

\begin{equation}
\begin{gathered}
\hat{F}_i = \text{Linear}(C_i, C)(F_i), \quad \forall i \\
\hat{F}_i = \text{Upsample}\left(\frac{W}{4} \times \frac{W}{4}\right)(\hat{F}_i), \quad \forall i \\
F = \text{Linear}(4C, C)(\text{Concat}(\hat{F}_i)), \quad \forall i \\
M = \text{Linear}(C, N_{\text{cls}})(F)
\end{gathered}
\end{equation}
where \( M \) refers to the predicted mask, and \(\text{Linear}(C_{\text{in}}, C_{\text{out}})\) refers to a linear layer with \( C_{\text{in}} \) and \( C_{\text{out}} \) as input and output vector dimensions respectively.

\subsection{Loss Function}
Utilizing Focal Loss to address class imbalance in LULC classification is an effective strategy, particularly in global LULC scenarios where classes such as forests and trees are disproportionately prevalent \cite{putty2024dual}. Focal Loss adapts the cross-entropy loss function to emphasize the learning of hard-to-classify or minority classes, enhancing model performance in scenarios characterized by significant class imbalances. In the paper,  \( \alpha\) -balanced variant is used and the expression is as follows:

\begin{equation}
F L\left(p_t\right)=-\alpha_{\mathrm{t}}\left(1-p_t\right)^\gamma \log \left(p_t\right)
\end{equation}
where
\begin{equation}
p_t=\left\{\begin{array}{lr}
p & \text { if } y=1 \\
1-p & \text { otherwise }
\end{array}\right.
\end{equation}
where \( p_t \) is the predicted probability of the class with the correct label, representing the model's confidence in its prediction. The parameter \( \alpha_t \) is a weighting factor used to balance the classes, typically set based on the inverse frequency of the classes to mitigate the effects of class imbalance. Besides, \( \gamma \) is the focusing parameter, which serves to decrease the loss contribution from easy examples, thereby prioritizing the learning from more difficult, misclassified examples.

\section{Experiment Setting}

In this section, we provide an overview of the benchmark models, experimental settings, and evaluation metrics used in our study.

\subsection{Benchmark Models}

In the paper, several well-established models have been selected as benchmarks to comprehensively evaluate the performance of the proposed model. The selection includes monomodal models such as PSPNet \cite{zhao2017pyramid}, DeepLabV3+ \cite{chen2018encoder}, CMTFNet \cite{wu2023cmtfnet}, Segformer \cite{xie2021segformer}, and ViT-Adapter \cite{chen2022vision}. Additionally,  bimodal models like MultiSenseSeg \cite{wang2024multisenseseg}, CMX \cite{zhang2023cmx}, FTransUNet \cite{ma2024multilevel} and CMNeXt \cite{zhang2023delivering} are also considered. To guarantee the integrity and fairness of the experimental evaluation, all models are trained from scratch. Moreover, in the visual analysis section, SOTA global LULC products from ESA and ESRI have been selected as benchmarks.

\subsection{Experiment Setting}
 
In the experiments, the input remote sensing imagery have dimensions of 510 × 510 pixels. The AdamW \cite{loshchilov2017decoupled} optimizer is chosen, and training is conducted over 200 epochs. A cosine annealing strategy is employed for dynamic learning rate adjustments, starting with an initial learning rate of 0.0005. The batch size is maximized at 4, within the memory constraints of the GPUs, and eight NVIDIA RTX 2080 Ti GPUs, each with 11 GB of memory, are used. To streamline the multi-GPU training process and ensure consistent batch sizes across different experiments, the Accelerator library is utilized. Finally, simple data augmentations, including random rotation and flipping, are applied in the paper.

\subsection{Evaluation Metrics}
In the paper, the network performance is evaluated from two aspects.
The first aspect focuses on network accuracy, assessed using metrics such as intersection over Union (IoU), mean IoU (mIoU), overall accuracy (OA), and F1-score. The second aspect evaluates the computational complexity and real-time performance of the network, which includes the number of parameters (Params), floating point operations Per Second (FLOPs), and frames per second (FPS). These dual aspects facilitate a comprehensive assessment of both the precision and computational efficiency of the network architectures under investigation.

\section{Experimental Results}
In this section, we present experimental results to verify the effectiveness of our proposed network for global LULC semantic segmentation in terms of both quantitative and qualitative evaluations. Besides, we conduct comprehensive ablation studies to evaluate the impact of various components, input modalities, branch parameter allocation strategies, and loss functions on segmentation performance.

\subsection{Results}

\afterpage{
\begin{figure*}[p] 
  \centering
  \includegraphics[scale=1]{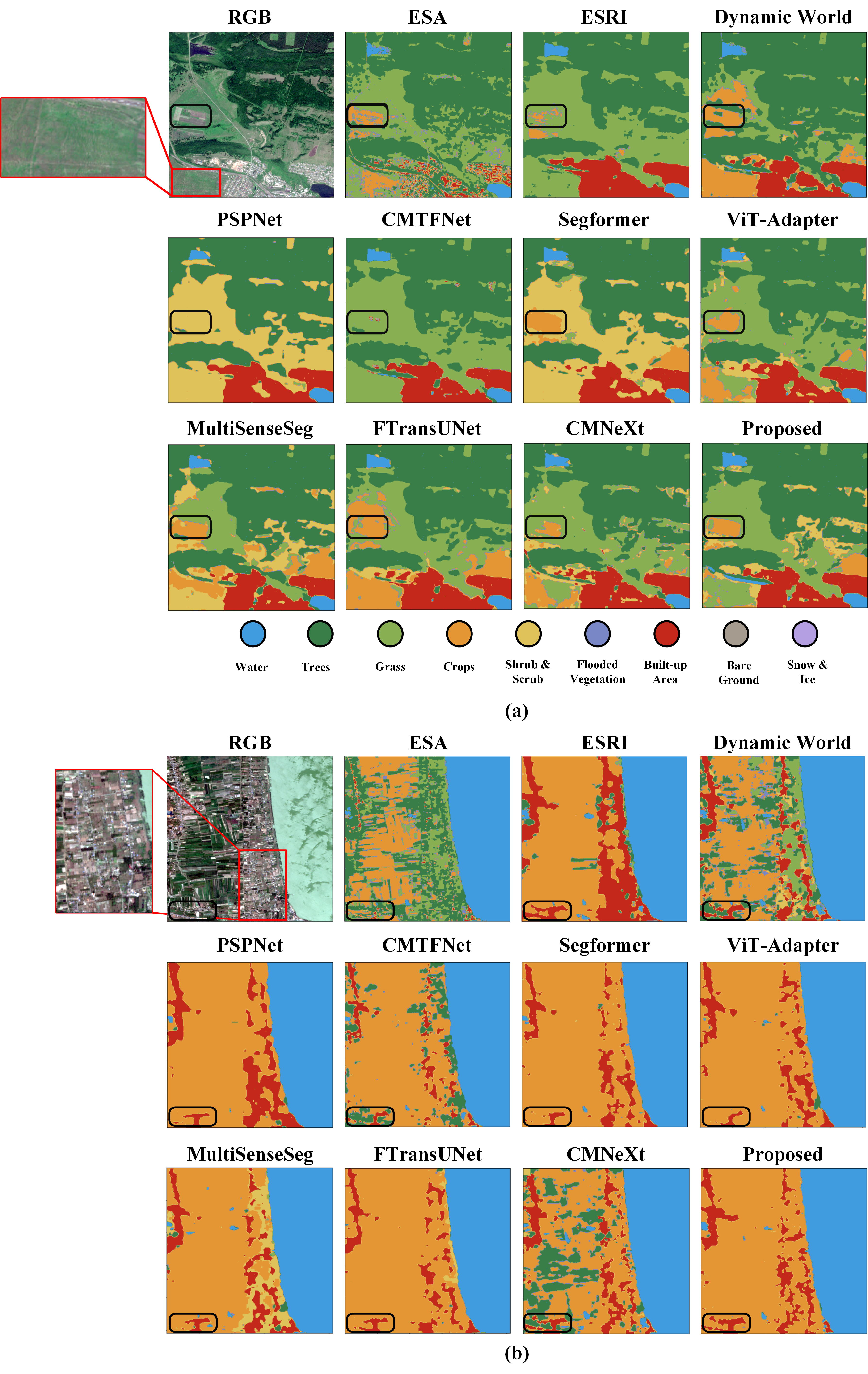} % 调整图像宽度以适应页面
  \caption{Visualization comparisons on the dataset. The area within the black bounding box emphasizes the performance disparities among the models, while the red box indicates the zoomed-in region. 
 \label{fig:VS_Results1}}
\end{figure*}
}
\subsubsection{Quantitative Analysis Results}
Tables \ref{tab: Evaluation Metrics network} and \ref{tab: PARAMETERS AND FLOPS} present the parameters, FLOPS, speed, and corresponding evaluation metrics for the various models compared in this study. The results demonstrate that within a specified range of parameters and FLOPS, the accuracy of our proposed network consistently exceeds that of other monomodal and bimodal models, both in terms of mIoU, OA and F1 scores. Notably, the proposed network achieves an mIoU of 59.5\%, an OA of 79.48\%, and an F1 score of 71.68\% with only 26.70M parameters, 109.59G FLOPS and 38.23 FPS. In comparison to the third-best method, CMX, our network shows an improvement of 3.76\% in mIoU, 2.41\% in OA, and 2.77\% in F1 scores. 

Additionally, Table \ref{table:class_results} displays the per-class results of different methods on the dataset for global LULC segmentation. It lists various methods and their performance in specific categories (such as water, trees, grass, built-up areas, etc.) using the percentage IoU metric. The proposed network demonstrates superior performance across multiple categories, with an IoU of 88.87\% in water, 46.55\% in crops, 74.90\% in shrub scrub, 78.92\% in built-up areas, and 80.08\% in snow/ice, achieving the highest IoU scores in these categories.

The network effectively integrates spectral and SAR features by a cross-attention mechanism, enhancing the specificity and efficiency of dual modalities information processing, which improves the model's performance. Moreover, this streamlined network design reduces the need for redundant parameters. 

%Furthermore, our method attained the highest scores across all evaluated categories. 
\subsubsection{Qualitative Analysis Results}
Fig. \ref{fig:VS_Results1} (a) and (b) display the representative visualization results for selected methods, highlighting the performance differences between the proposed network and other SOTA deep learning methods. It is observed that the classification map generated by the official Dynamic World exhibits significant fragmentation, leading to ambiguous boundaries. In contrast, the classification map from the proposed method more accurately reflects the actual land cover distribution. 

The proposed network delivers highly effective classification maps that excel in texture and edge detail compared to other deep-learning approaches. This superior performance is evident in the clear class boundaries and enhanced visual quality of the maps, which are achieved by integrating additional modalities during the feature extraction and learning phases. By improving the exchange of information between MSI and SAR modalities, the network enhances its ability to process and combine complementary data. This capability enables the network to uncover robust non-linear sequential relationships within the remote sensing data, resulting in more realistic and detailed depictions of LULC regions in the classification maps.

\begin{table}[!t]
\renewcommand{\arraystretch}{1.3}
\centering
\caption{Evaluation Metrics on the Dynamic World+ Dataset
\label{tab: Evaluation Metrics network}}
\footnotesize % 设置表格字体大小为footnotesize，约等于8号字体
% \resizebox{0.8\columnwidth}

% {!}
{%

\begin{tabular}{cccc}
\hline \hline \\[-3mm]

\multicolumn{1}{c}{Methods} & \multicolumn{1}{c}{mIoU(\%)}
& \multicolumn{1}{c}{OA(\%)}
& \multicolumn{1}{c}{F1(\%)}

\\[-3mm]\\\hline \\[-3mm]

Dynamic World & 52.22 & 72.86 & 65.74\\

PSPNet & 54.57 & 76.93 & 66.41\\

DeepLabV3+ & 54.78 & 77.33  & 66.04 \\

CMTFNet & 52.80 & 74.10 & 65.33\\

Segformer & 58.33 & 79.06  & 70.34\\

ViT-Adapter & 55.75 & 76.44 & 68.16\\
[-3mm]\\\hline \\[-3mm]

MultiSenseSeg & 48.89 & 70.41 & 61.22\\

CMX & 57.42 & 77.61 & 69.75\\

FTransUNet & 54.85 & 77.21 & 66.96\\

CMNeXt & 56.61 & 75.85 & 69.42\\

\rowcolor{gray!20} Proposed & \textbf{59.58} &  \textbf{79.48}  &  \textbf{71.68}\\
\\ [-3mm]\hline \hline \\[-3mm]

\end{tabular}
}%
\end{table}

Specially, an analysis of the classification maps in Fig. \ref{fig:VS_Results1} (a) reveals that the proposed method significantly reduces the number of misclassified pixels, and excels in maintaining classification coherence and delivering clear boundary accuracy. Within the area delineated by the black frame, the proposed classification map demonstrates precise boundary delineation and effectively minimizes the misclassification of vegetative types, particularly distinguishing crops from other types of vegetation. Despite the satellite imagery clearly depicting this region as farmland, several other methods, such as PSPNet and CMTFNet, tend to misclassify it, failing to recognize its agricultural nature. Additionally, in the detailed view provided by the red zoomed-in section, many networks struggle with accurately classifying land cover, misidentifying obvious grassland as crop or shrub and scrub. 

 Moreover, in Fig. \ref{fig:VS_Results1} (b), within the area delineated by the black frame, only ESRI and the proposed network successfully reveal a coherent built-up area, distinguishing themselves from other methods. However, in the zoomed-in section of the ESRI map, the depiction of the built-up area is overly extensive, incorrectly classifying some pixels that should be identified as cropland. This discrepancy highlights the superior performance of the proposed network in accurately distinguishing between built-up and agricultural areas. Within the red zoomed-in region, many networks, except for the proposed network, exhibit significant challenges in distinguishing between urban areas and farmland, failing to effectively recognize and differentiate these two types of land cover.

% \begin{figure*}[b]
%   \centering

%     \includegraphics[scale=1]{Figures/VS_Results2.png}
%     \caption{ Visualization comparisons on the dataset. \label{fig:VS_Results2}}
% \end{figure*}

\begin{table}[!t]
\renewcommand{\arraystretch}{1.3}
\centering
\footnotesize
\caption{Parameters, FLOPS and Speed
\label{tab: PARAMETERS AND FLOPS}}
\footnotesize % 设置表格字体大小为footnotesize，约等于8号字体
% \resizebox{0.8\columnwidth}{!}
{%

\begin{tabular}{cccc}
\hline \hline \\[-3mm]

\multicolumn{1}{c}{\textbf{Methods}} & \multicolumn{1}{c}{\textbf{Params (M)}}
& \multicolumn{1}{c}{\textbf{FLOPS (G)}}
& \multicolumn{1}{c}{\textbf{Speed (FPS)}}

\\[-3mm]\\\hline \\[-3mm]

PSPNet & 68.07 & 545.54 & 28.09\\

DeepLabV3+ & 57.56 & 174.42 & 28.79\\

CMTFNet & 30.10 & 34.93 & \textbf{42.26}\\

Segformer & 27.38 & 114.53 & 36.92\\

ViT-Adapter & 179.35 & 1978.31 & 23.98\\
[-3mm]\\\hline \\[-3mm]

MultiSenseSeg & 58.62 & 531.50 & 25.42\\

CMX & 66.58 & 114.74 & 36.29\\

FTransUNet & 184.14 & 171.13 & 37.08\\

CMNeXt & 58.69 & 117.09 & 36.62 \\

%Proposed-half & 21.77 & 103.58\\

\rowcolor{gray!20} Proposed & \textbf{26.70} &  \textbf{109.59} & 38.23\\

\\ [-3mm]\hline \hline \\[-3mm]

\end{tabular}
}%
\end{table}

\begin{table*}[t]
\centering
\caption{Per-Class Results on the Dataset for Global LULC Semantic Segmentation}
\label{table:class_results}
\begin{tabular}{|c|c|c|c|c|c|c|c|c|c|c|}
\hline
\textbf{Method} & \textbf{Water} & \textbf{Trees} & \textbf{Grass} & \textbf{Crops} & \makecell{\textbf{Shrub} \\ \textbf{scrub}} & 
\makecell{\textbf{Flooded} \\ \textbf{Vegetation}}
 & \textbf{Built-up Area} & \textbf{Bare Ground} & \makecell{\textbf{Snow} \\ \textbf{Ice}}

 &  \textbf{mIoU} \\
\hline
PSPNet  & 83.47 & 75.80 & 13.64 & 35.14 & 68.37 & 39.48 & 79.09 & 18.02 & 78.14 & 54.57  \\
DeepLabV3+  & 83.38 & 74.06 & 7.87 & 39.17 & 73.59 & 40.04 & 79.11 & 16.87 & 78.95 & 54.78  \\
CMTFNet  & 81.03 & 68.98 & 18.51 & 26.99 & 67.18 & 33.50 & 78.56 & 22.68 & 77.79 & 52.80  \\
Segformer  & 87.16 & 75.10 & \textbf{27.07} & \textbf{49.39} & 72.54 & 39.65 & 79.05 & 16.49 & 78.48 & 58.33  \\
ViT-Adapter  & 83.76 & 72.01 & 16.37 & 38.33 & 71.75 & 36.71 & 76.97 & 27.00 &   78.88 & 55.75 \\
\hline
MultiSenseSeg  & 86.40 & 72.83 & 14.49 & 24.42 & 57.22 & 30.79 & 63.20 & 14.87 & 75.83 & 48.89  \\
CMX & 83.90 & 71.44 & 20.13 & 39.74 & 74.80 &  37.33 & \textbf{81.38} & 28.69 & 79.41 & 57.42  \\
FTransUNet & 85.71 & 73.60 & 20.10  & 31.01 & 72.26 & 39.69 & 77.94 & 17.93 & 75.39 & 54.85 \\
CMNeXt & 88.35 & 72.44 & 19.92 & 43.96 & 64.42 & 37.04 & 74.97 & \textbf{31.03} & 77.34 & 56.61  \\
\hline
\rowcolor{gray!20} Proposed (MiT-B2) & \textbf {88.87} & \textbf{75.85} & 25.10 &  46.55 & \textbf{74.90}  & \textbf{40.36} & 78.92 & 25.64 & \textbf{80.08} & \textbf{59.58}  \\

\hline
\end{tabular}
\end{table*}

%\subsection{European Country Case}

\subsection{Ablation Study}

To thoroughly assess the contribution of each component module, we systematically remove the corresponding modules from the designed network. Additionally, ablation experiments are conducted on input modalities, branch parameter allocation, and loss function selection.

\begin{table}[!t]
\renewcommand{\arraystretch}{1.3}
\centering
\caption{Evaluation Metrics for Component Ablation Experiment}
\label{tab: Evaluation Metrics for Module Ablation}

\fontsize{8}{10}\selectfont % 设置字体大小为8pt，不显式设置行距
% \resizebox{0.9\columnwidth}{!}

{%

\begin{tabular}{ccc||ccc}
\hline \hline \\[-3mm]

\multicolumn{1}{c}{CA} 
& \multicolumn{1}{c}{Eff. SA} 
& \multicolumn{1}{c}{MMAM}
& \multicolumn{1}{c}{mIoU(\%)}
& \multicolumn{1}{c}{OA(\%)}
& \multicolumn{1}{c}{F1(\%)}

\\[-3mm]\\\hline \\[-3mm]

{} & {} & {} & 55.91 & 77.35 & 67.88\\
\checkmark & {} & {} & 57.13 & 77.87 &69.60\\
{} & \checkmark & {} & 57.16 & 76.05 &70.09\\
{} & {} & \checkmark & 57.23 & 78.31 & 69.32\\
{} & \checkmark & \checkmark & 57.56 & 77.55 & 69.67\\
\checkmark & {} & \checkmark & 57.51 & 77.30  & 70.05 \\
\checkmark & \checkmark & {} & 57.83 & 78.89 & 69.80\\

\rowcolor{gray!20} \checkmark & \checkmark & \checkmark & \textbf{59.58} & \textbf{79.48}  & \textbf{71.68} \\

\\ [-3mm]\hline \hline \\[-3mm]

\end{tabular}
}%
\end{table}

\subsubsection{Components}
To verify the effectiveness of each component and the corresponding combinations proposed in the network, ablation studies are conducted by progressively removing specific modules or their combinations, while retaining the dual-branch framework. As detailed in Table \ref{tab: Evaluation Metrics for Module Ablation}, seven distinct ablation experiments are implemented based on our network architecture.

In the initial experiment, no modules are included, resulting in the lowest mIoU score among all the experiments. The subsequent experiments two, three, and four retain only the Multispectral-to-SAR cross attention and SAR-to-multispectral cross attention modules, efficient self-attention, and MMAM, respectively. These experiments demonstrate that the inclusion of each module individually enhances the accuracy of global LULC segmentation compared to the baseline without any modules. In the fifth experiment, the Multispectral-to-SAR cross attention and SAR-to-multispectral cross attention module are selectively removed from the trio of modules to assess their impact on performance. The sixth experiment involves omitting the efficient Self-Attention module and evaluating its contribution to overall accuracy. In the seventh experiment, the MMAM is replaced with a simple concatenation operation to test the efficacy of the advanced fusion strategy compared to basic concatenation for feature integration. The outcomes of these experiments indicate a decrease in performance metrics relative to the full module configuration, yet there is a discernible improvement compared to scenarios where only a single module is used. These results confirm the critical role of each individual module and their combinations in enhancing the functionality and accuracy of the network architecture.

\subsubsection{Input modalities}
We conducte band ablation studies using the single-branch structure of our designed network. According to the results shown in Table \ref{tab: Evaluation Metrics for band Ablation}, the integration of multimodal and multiband data significantly enhances model performance. As the number of modalities and bands increases, performance metrics progressively improve, reaching a maximum mIoU of 58.33\%, OA of 79.06\%, and an F1 score of 70.34\%. Comparative experiments using solely SAR data revealed that SAR data alone is inadequate for the semantic segmentation task due to its limited spatial information. Given its sensitivity to the geometric and structural attributes of targets, SAR data is better suited as a complementary perspective to multispectral information, enhancing the network's predictive accuracy. Additionally, the results from Table \ref{tab: Evaluation Metrics for band Ablation}, which pertain to the use of only multispectral and SAR data in a single-branch configuration, indicate that the single-branch structure was ineffective in facilitating meaningful interaction between SAR and multispectral data, preventing the model from leveraging the potential synergistic advantages of these two data sources. This highlights the superiority of our designed network, which utilizes a dual-branch structure with cross-attention mechanisms to interact modalities effectively.

Regarding the dual-branch band ablation experiments, the experimental outcomes with RGB and other bands underscore the necessity of meticulous division of branches when inputting data into the network. This demonstrates the significance of precisely configuring the network architecture when processing different types of remote sensing data, ensuring that each data type can fully leverage its unique advantages.

\begin{table}[!t]
\renewcommand{\arraystretch}{1.3}
\centering
\caption{Evaluation Metrics for Band Ablation Experiments}
\label{tab: Evaluation Metrics for band Ablation}
\footnotesize % 设置表格字体大小为footnotesize，约等于8号字体
% \resizebox{0.9\columnwidth}{!}

{%

\begin{tabular}{cccc}
\hline \hline \\[-3mm]

\multicolumn{1}{c}{Input bands} & \multicolumn{1}{c}{mIoU(\%)}
& \multicolumn{1}{c}{OA(\%)}
& \multicolumn{1}{c}{F1(\%)}

\\[-3mm]\\\hline \\[-3mm]
%Segformer & 58.33 & 79.06  & 70.34\\
% RGB & 56.68 & 78.32 & 68.43\\

% {\parbox{3cm}{\centering SAR \\ (2 bands, Mono-) }}
{\parbox{3cm}{\centering SAR \\ (2 bands, Mono-) }} & 23.61 & 51.07 & 34.90\\
{\parbox{3cm}{\centering \vspace{0.2cm} RGB \\ (3 bands, Mono-) }} & 56.58 & 78.46 & 68.19\\
{\parbox{3cm}{\centering \vspace{0.2cm} Multispectral \\ (10 bands, Mono-) }} & 58.04 & 78.44 & 70.19\\

{\parbox{3cm}{\centering \vspace{0.2cm} Multispectral + SAR \\ (12 bands, Mono-) }} & 58.33 & 79.06  & 70.34  \vspace{0.1cm} \\

\hline \\[-3mm]
{\parbox{3cm}{\centering RGB + Other \\ (3 bands + 9 bands) }} & 57.51 & 77.87  & 69.78\\
\rowcolor{gray!20}
{\parbox{3cm}{\centering  \vspace{0.2cm} Multispectral + SAR \\ (12 bands, Dual-) }} & \textbf{59.58} &  \textbf{79.48}  &  \textbf{71.68}\\

% Proposed-half & 58.6 & 78.6 & 70.9\\
\\ [-3mm]\hline \hline \\[-3mm]

\end{tabular}
}%
\end{table}

\subsubsection{Branch parameter allocation strategy}

Table \ref{tab: Evaluation Metrics for unbalanced parameters} presents the evaluation metrics for the different branch parameter allocation strategies. It is evident from the results that, without altering the network architecture, the evaluation metrics decrease as the parameter configuration for the SAR branch increases. This finding suggests that parameter allocations should be strategically adjusted based on the importance of the branch modality for LULC semantic segmentation. Modality branches of greater importance should be allocated relatively more parameters to optimize performance.

\subsubsection{Loss functions}
\begin{figure}[t]
  \centering
    \includegraphics[scale=1]{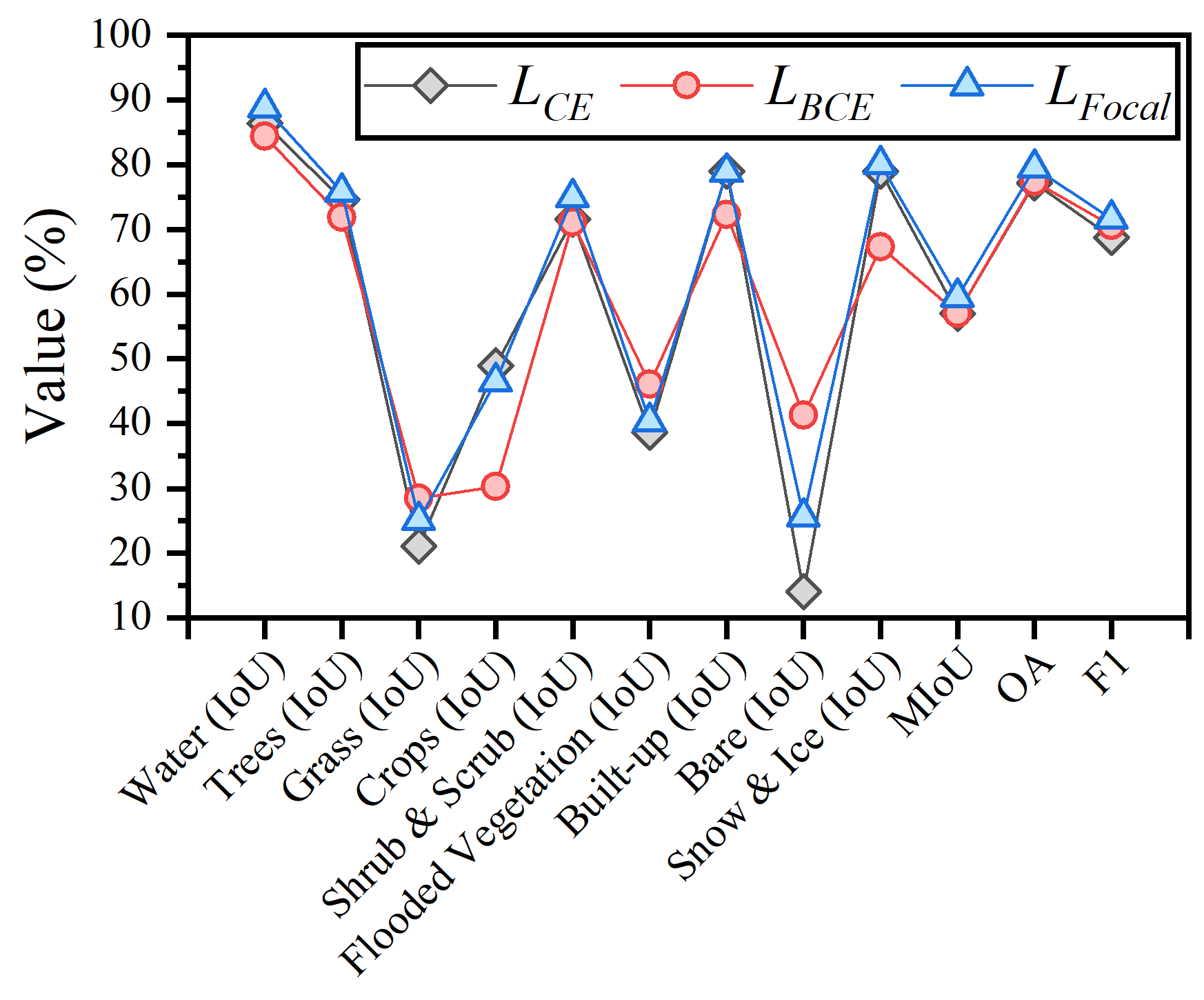}
    \caption{ Per-class results and evaluation metrics for different loss functions.
 \label{fig:Different_loss}}
\end{figure}

The ablation study results on the effect of different loss functions are shown in Fig. \ref{fig:Different_loss}. As depicted in the figure, the experimental results are clearly visualized through lines and numeric values. 
$L_{Focal}$ consistently shows superior performance compared to $L_{BCE}$ and $L_{CE}$ across most categories. This suggests that $L_{Focal}$ is more effective in addressing class imbalance by decreasing the weight of easily classified samples and increasing the weight for hard-to-classify ones, thereby enhancing the model's recognition ability for minority classes. Comparatively, $L_{BCE}$ also generally performs better than $L_{CE}$, indicating that it is somewhat effective in handling class imbalance but not as potent as $L_{Focal}$.

\begin{table}[!t]
\renewcommand{\arraystretch}{1.3}
\centering
\caption{Evaluation Metrics for Branch Parameter Allocation Strategy}
\label{tab: Evaluation Metrics for unbalanced parameters}
\footnotesize % 设置表格字体大小为footnotesize，约等于8号字体
% \resizebox{0.75\columnwidth}{!}

{%

\begin{tabular}{cccc}
\hline \hline \\[-3mm]

\multicolumn{1}{c}{\parbox{3cm}{\centering Imbalance ratio of \\ parameters across \\ different branches \\ (Multispectral, SAR) }}

& \multicolumn{1}{c}{mIoU(\%)}
& \multicolumn{1}{c}{OA(\%)}
& \multicolumn{1}{c}{F1(\%)}

\\[-3mm]\\\hline \\[-3mm]

% 2/3 1/3 & - & - & - & - & -\\
1/4, 3/4 & 55.36 & 76.73 & 67.37 \\
1/2, 1/2 & 58.26 & 77.55 & 71.03 \\
\rowcolor{gray!20} 3/4, 1/4 (Proposed) & 59.58 & 79.48 & 71.68 \\

% Proposed-half & 58.6 & 78.6 & 70.9\\
\\ [-3mm]\hline \hline \\[-3mm]

\end{tabular}
}%
\end{table}

% \section{Discussion}
\section{Conclusion}
In this paper, we first expanded the SAR data and then aligned it with the multispectral data within the original Dynamic World dataset, introducing the \dName dataset and validating its practicality. Subsequently, we proposed a transformer-based multimodal network, named \nName, which was designed to effectively handle the practical task of global land use and land cover (LULC) classification with as few parameters and FLOPs as possible. We introduced modules such as the \crossAttentionModule{} with a differentiated parameter configuration strategy and the \FusionModule{}, which effectively facilitated both the interaction and fusion of multiscale features extracted by different modal encoders. The combination of these modules enabled the model to more accurately capture essential features when processing multimodal inputs, significantly enhancing the effectiveness of the LULC semantic segmentation task. Overall, the advantages of the proposed \nName not only lay in its streamlined structure but also in its state-of-the-art performance, which was demonstrated through comparative experiments and extensive ablation studies.
In the future, the research team plans to continue developing more interpretable and reliable multimodal deep learning models aimed at effectively handling vulnerable incomplete modal Earth observation data and enhancing the accuracy and response speed of remote sensing image segmentation in complex geospatial semantic contexts.

\bibliographystyle{IEEEtranTIE}
\bibliography{Reference.bib}\ %IEEEabrv instead of IEEEfull

\newpage

\end{document}